\documentclass[structabstract]{aa}  
\usepackage{graphicx}
\usepackage{txfonts}
\usepackage[authoryear]{natbib}
\newcommand{\src}  {SWIFT J1626.6--5156}
\def\simless{\mathbin{\lower 3pt\hbox
     {$\rlap{\raise 5pt\hbox{$\char'074$}}\mathchar"7218$}}}   
\def\simmore{\mathbin{\lower 3pt\hbox
     {$\rlap{\raise 5pt\hbox{$\char'076$}}\mathchar"7218$}}}   

\begin{document}
   \title{Multi-frequency observations of SWIFT J1626.6-5156}

   \subtitle{}

   \author{P. Reig
          \inst{1,2}
          \and
          E. Nespoli\inst{3,4}    
	  \and
	  J. Fabregat \inst{3}
	  \and
	  R.E. Mennickent \inst{5}
          }

   \institute{IESL, Foundation for Research \& Technology-Hellas
              GR-71110, Heraklion, Crete, Greece\\
              \email{pau@physics.uoc.gr}
         \and
             Department of Physics and Institute of Theoretical \& Computational
	     Physics, University of Crete, GR-71003 Heraklion, Greece\\
             \email{}
	\and
	Observatorio Astron\'omico Universidad de Valencia, 
	C/ Catedr\'atico Agust\'{\i}n Escardino Benlloch 7, 46980 Paterna,
	Valencia, Spain\\  
	\email{elisa.nespoli@uv.es, juan.fabregat@uv.es} 
	\and
	Valencian International University, Pr. C/ Jos\'e Pradas Gallen, 
	12006 Castell\'on de la Plana, Spain \\
          \email{}
	\and
	Departamento de Astronom\'{\i}a, Universidad de Concepci\'on, Chile, 
	Casilla 160-C, Concepci\'on, Chile \\
          \email{rmennick@udec.cl}
	}

   \date{}

 
  \abstract
   {\src\ is an X-ray pulsar that was discovered in
   December 2005 during an X-ray outburst. Although the X-ray data suggest
   that the system is a high-mass X-ray binary, very little information exists 
   on the nature of the optical counterpart. }
   {We investigate the emission properties of the optical
   counterpart in the optical and near-IR bands and the
   long-term  X-ray variability of the system in order 
   to determine unambiguously the nature of this X-ray pulsar.}
   {We have performed an X/optical/IR analysis of \src. We have analysed all 
   RXTE observations since its discovery, archived optical
   spectroscopic and photometric data and obtained for the first time near-IR
   spectra. X-ray energy  spectra
   were fitted with  models composed by a combination of photoelectric absorption, a
power law with high-energy exponential cutoff and a Gaussian line profile
at  6.5 keV and an absorption edge at around 9 keV. X-ray power spectra were
fitted with Lorentzian profiles. We identified and measured the equivalent
width and relative intensity of the spectral features in the optical and 
infrared spectra to determine the spectral type of the optical counterpart. }
   {
The K-band spectrum shows He\,I $\lambda$20581\AA\ and H\,I $\lambda$21660 \AA\  (Brackett-gamma)
in emission, which confine the spectral type of the companion to be earlier
than B2.5. The H-band spectrum exhibits the HI Br-18-11 recombination 
series  in emission. The most prominent feature of the optical band spectrum
is the strong emission of the Balmer line H$\alpha$. The 4000-5000 \AA\ 
spectrum contains He\,II and numerous He\,I lines in absorption, indicating 
an early B-type star. The
source shows three consecutive stages characterised by  different types of 
variability in the X-ray band:  a
smooth decay after the peak of a large outburst, large-amplitude flaring 
variability (reminiscent of type I oytbursts) and quiescence. We observed that the spectrum becomes softer as
the flux decreases and that this is a common characteristic of the X-ray
emission for all observing epochs. An emission line feature at $\sim$6.5 keV is
also always present. 
      }
   {
The X-ray/optical/IR continuum and spectral features are typical of an 
accreting X-ray pulsar with an early-type donor.
The long-term X-ray variability exhibiting large outbursts, 
minor outbursts and quiescent emission is also characteristic of hard X-ray
transients. We conclude that \src\ is a Be/X-ray binary with a B0Ve
companion located at a distance of $\sim$10 kpc.
}


   \maketitle
%

\section{Introduction}

\src\ is a hard X-ray transient whose actual nature is uncertain. It was
discovered on 18 December 2005 by {\em Swift}/BAT when its X-ray emission
was showing short-term flaring episodes \citep{palmer05}. Immediate after
the first detection by {\em Swift}, the source was observed by {\em RXTE},
which confirmed the presence of an X-ray pulsar with $P_{\rm spin}=15.377$
s and strong variations of the pulse fraction during the flares
\citep{markwardt05, belloni06}. In a subsequent observation with {\em
Swift}/XRT, the X-ray position was refined to RA (J2000): 16:26:36.24, DEC
(J2000): --51:56:33.5, with a 90\% error radius of 3.5'' \citep{campana06}.
The optical and infrared observations of the only counterpart so far
proposed are contradictory. While optical spectroscopy suggests a Be star
\citep{negueruela06}, infrared observations seem to indicate a late-type
object \citep{rea06}. The $\gamma$-ray mission {\em INTEGRAL} also detected
\src\ about a year after its discovery, but no X-ray flares were observed
\citep{tarana06}.

The first detailed study of \src\ was carried out by \citet{reig08}, who
found that the duration of the flares was a few hundred seconds and that
the X-ray intensity increased by a factor of 3.5 during the flares. They
also presented evidence (although not conclusive) in favour of a high-mass
X-ray binary classification of the system. The determination of the nature
of the system is very important because the flares seen in \src\ would
constitute the shortest events of this kind ever reported in a high-mass
X-ray binary.

In a recent paper, \citet{baykal10}(see also \citealt{icdem11}) performed a
pulse frequency analysis and reported the discovery of the orbital
parameters of the system. They also studied the evolution of the neutron
star spin period throughout the outburst. The periodic trend of pulse
frequencies yielded an orbital period of $P_{\rm orb}=132.89\pm0.03$ days
and an eccentricity of $e=0.08\pm0.01$. Pulse-phase spectroscopy has also
been addressed by \citet{reig08}, while a preliminary analysis of the
evolution of the pulse profiles can be found in \citet{baykal11}. Here we
present a detailed X-ray spectral and timing analysis of the entire 2006 outburst and
subsequent flaring emission. The X-ray varibility that followed the
major outburst is reminiscent of the so-called type-I outburst, seen in
many Be/X-ray binaries \citep[see e.g.][]{wilson02}. However, unlike
typical Be/X-ray binaries, the frequency of the outbursts observed in
\src\, does not always equal the orbital period \citep{baykal10}. For this
reason, we shall refer to these quasiperiodic enhancements of the X-ray
intensity as flares (not to be confused with the three short-term flares
reported by \citet{reig08}).  We also analysed data during the X-ray quiescent
state that followed the flaring state and present new optical and infrared 
observations. Our aim is to solve the nature of this unusual accreting
X-ray pulsar.

   \begin{figure}
   \centering
   \includegraphics[width=8cm]{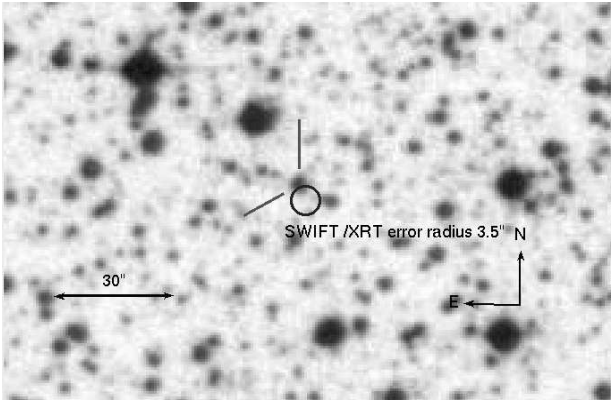}
   \caption{Finding chart with the identification of the optical
   counterpart to \src. The image was downloaded from the Aladin Sky Atlas.}
              \label{chart}
    \end{figure}
   \begin{table}
   \begin{center}
      \caption[]{Optical/infrared apparent magnitudes of the 
      optical counterpart to \src\ from the NOMAD catalogue. Note that
      $B$, $V$, $R$ and $I$ are photographic magnitudes.}
         \label{cat} 
         \begin{tabular}{ccccccc}
            \hline  \noalign{\smallskip}
B	&V	&R	&I	&J	&H	&K \\
            \noalign{\smallskip}  \hline
            \noalign{\smallskip}
16.81	&15.54	&15.81	&14.27	&13.44	&12.95	&12.54 \\
            \noalign{\smallskip}
            \hline
         \end{tabular}
	\end{center}
   \end{table}

\begin{table}[!ht]
\caption{$H$ and $K$-band line identifications and measured equivalent
widths for 2MASS J16263652--5156305.}
\begin{center}
\begin{tabular}{lcrr}
\hline
\hline
  \noalign{\smallskip}
Spectral	&  Wavelength   & Equivalent    \\ 
feature		&($\mu$m)	&width (\AA)	\\
   \noalign{\smallskip}
\hline 
  \noalign{\smallskip}
Br-10		&1.737	&--7.8$\pm$0.3  \\
Br-11		&1.678 	&--7.9$\pm$0.5  \\
Br-12		&1.641 	&--8.3$\pm$0.5  \\
Br-13		&1.611 	&--7.0$\pm$0.5  \\
Br-14		&1.588 	&--9.7$\pm$0.6  \\
Br-15		&1.570 	&--8.3$\pm$0.5  \\
Br-16		&1.556	&--7.8$\pm$0.2  \\
Br-17		&1.544  &--8.0$\pm$0.3  \\
Br-18		&1.535  &--6.2$\pm$0.3  \\
\ion{He}{I}	&2.058  &--9.9$\pm$0.4  \\ 
Br$ \gamma$	&2.166	&--8.8$\pm$0.3  \\
  \noalign{\smallskip}
  \hline
\end{tabular}
\end{center}
\label{ewir}
\end{table}

   \begin{figure}
   \centering
   \begin{tabular}{c}
   \includegraphics[width=8cm]{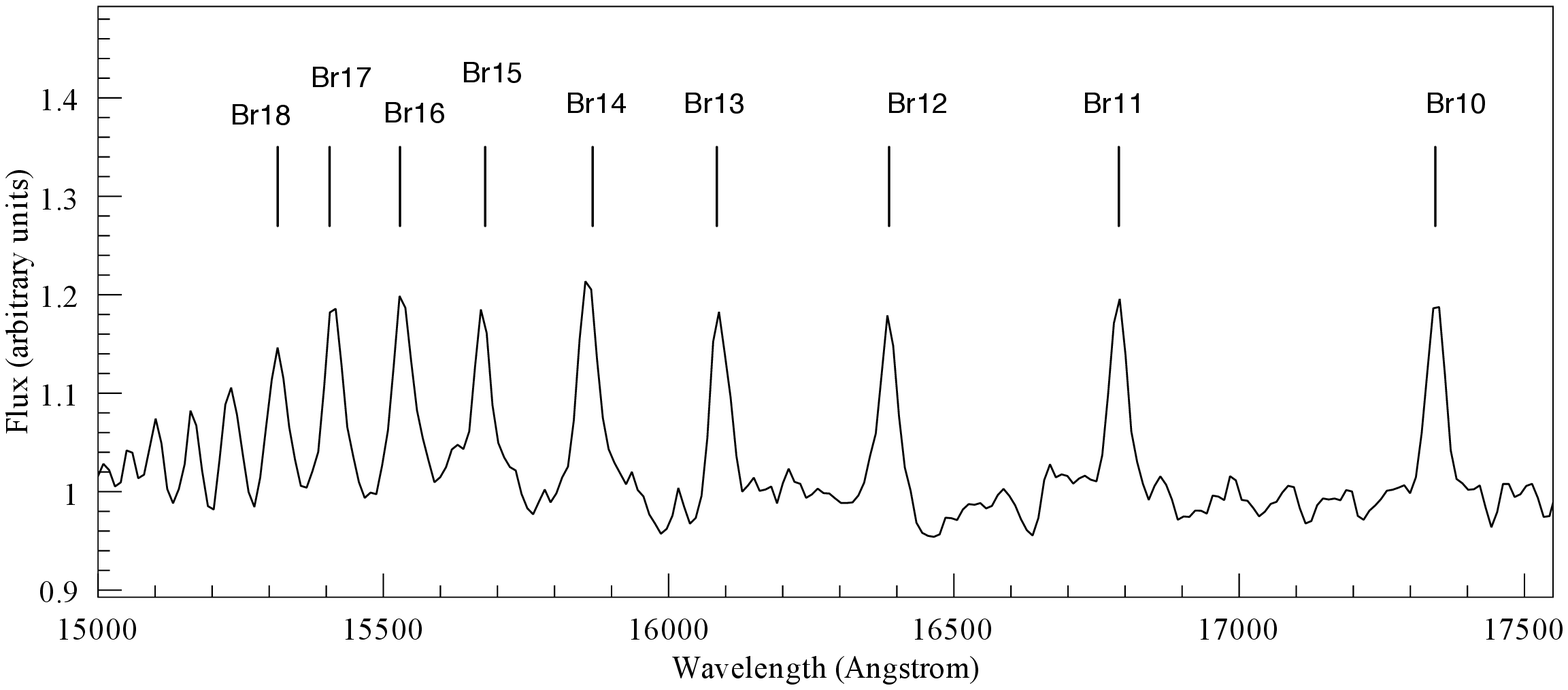} \\
   \includegraphics[width=8cm]{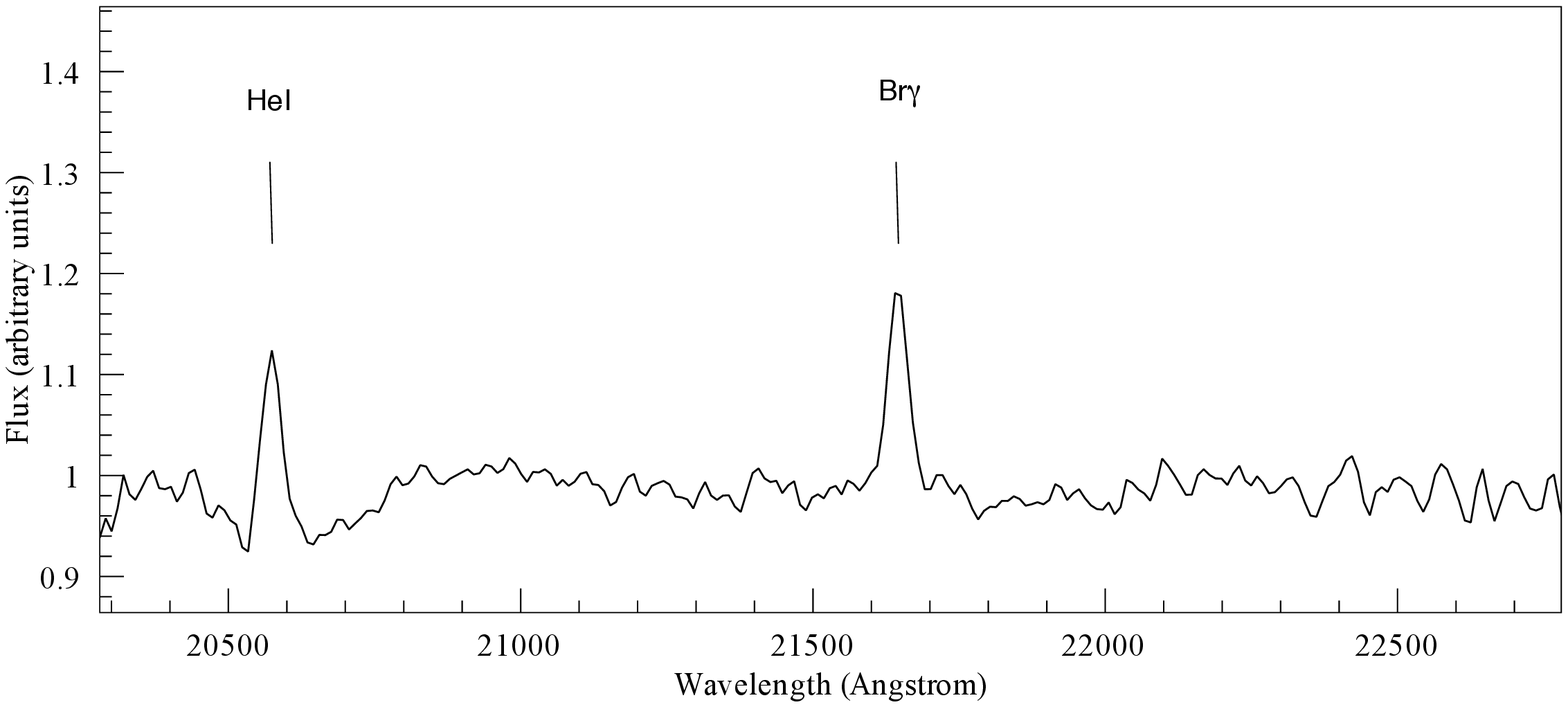} \\
   \end{tabular}
      \caption{H- and K-band spectra of the infrared counterpart to \src. 
      Helium and hydrogen lines are clearly seen in emission.}
         \label{irspec}
   \end{figure}

\begin{table*}
\caption{Log of the optical spectroscopic observations}
\label{optobs}
\begin{center}
\begin{tabular}{lcllccc}
\hline
\hline
  \noalign{\smallskip}
Date	&MJD	&instrument	&Grism	&Wavelength	&EW(H$\alpha$)	&EW(H$\beta$)    \\ 
	&	&set-up		&number	&range(\AA)	&(\AA)		&(\AA)\\
   \noalign{\smallskip}
\hline 
15-02-2006	&53782.359 &EMMI/NTT	&grism\#6  &5700--8700	&$-40\pm2$	&--	\\
17-09-2007	&54361.008 &EFOSC/3P6	&grism\#18 &4700--6770	&$-45\pm2$	&$-5.0\pm0.9$	\\
17-09-2007	&54360.984 &EFOSC/3P6	&grism\#14 &3095--5085	&--		&$-4.3\pm0.5$	\\
18-09-2007	&54362.025 &EFOSC/3P6	&grism\#14 &3095--5085	&--		&$-4.2\pm0.5^a$	\\
18-09-2007	&54361.993 &EFOSC/3P6	&grism\#3  &3050--6100	&--		&$-4.5\pm0.5^b$	\\
01-06-2009	&54983.078 &EFOSC/NTT	&grism\#13 &3685--9315	&$-45\pm2$	&$-5.1\pm0.9$	\\
  \noalign{\smallskip}
  \hline
\multicolumn{7}{l}{$a$: Average of three measurements} \\
\multicolumn{7}{l}{$b$: Average of two measurements} \\
\end{tabular}
\end{center}
\end{table*}

\section{Optical and infrared observations}

\subsection{NIR Observations}

Figure~\ref{chart} shows a digital image of the field around the X-ray best
position with the identification of the optical counterpart. Near-IR data
were obtained in visiting mode in July 2010, at the European Southern
Observatory (ESO). The employed instrument was the SofI spectrograph
\citep{moorwood98}, on the 3.5m New Technology Telescope (NTT) at La Silla,
Chile.  We used the long slit spectroscopy mode, at low resolution ($R =
588$) with the 1.53--2.52 $\mu$m  grism  and  1\arcsec\ width slit. The
instrument large field objective provided a FOV of $4.92'  \  \times \
4.92'$. The sky had thin cirri and seeing averaged between $0.9''$ and
$1.3''$.

With the aim of ensuring accurate removal of atmospheric features from the
spectrum, we followed a strategy similar to that outlined by
\citet{clark00}. At the telescope, we observed an \mbox{A0IV} standard
star immediately before the target and a G3 V immediately after it, in
order to obtain very small differences in airmass (differences between 0.02
and 0.04 airmasses were accomplished). To compute the telluric features in
the region of the H\,I 21\,661 \AA\ (Brackett-$\gamma$ line, or
Br$\gamma$), which is the only non-telluric feature in the A-star spectra,
we employed the observed G-star spectra divided by the solar
spectrum\footnote{We used the NSO/Kitt Peak FTS solar spectrum, produced by
NSF/NOAO.} properly degraded in resolution. The dispersion solution obtained
for the SofI spectra was also applied and the spectra of the A star, G star
and the solar one were aligned in wavelength space. For the $K$ spectral
region, a telluric spectrum was obtained by patching into the A-star
spectrum the ratio between the G star and the solar spectrum in the
Br$\gamma$ region (we selected the range 21\,590 - 21\,739 \AA). For the
$H$ spectral region, we employed as the telluric spectrum, the ratio
between the G-star and the solar spectrum.

Data reduction was performed using the IRAF\footnote{IRAF is distributed by
the National Optical Astronomy Observatories which is operated by the
Association of Universities for Research in Astronomy, Inc. under contract
with the National Science Foundation.} package, following the standard
procedure. We first corrected for the inter-quadrant row cross-talk, a
feature that affects the SofI detector; we then applied sky subtraction; we
employed dome flat-fields and extracted the one dimensional spectrum.
Wavelength calibration was accomplished using Xenon and Neon lamp spectra.
Spurious features, such as cosmic rays or bad pixels, were removed by
interpolation, when necessary. The reduced spectra, one covering the $H$
band, one the $K$ band, were normalized by dividing them by a fitted
polynomial continuum. We finally corrected for telluric absorption,
dividing each scientific spectrum by its corresponding telluric spectrum,
obtained as described above. A scale and a shift factor were applied to the
telluric spectrum, to best correct for the airmass difference and the
possible wavelength shift; the optimum values for these parameters were
obtained using an iterative procedure that minimizes the residual noise.
The final $H$ and $K$ spectra are shown in Fig.~\ref{irspec}.

\subsection{Optical spectra}

Optical spectroscopic data were retrieved from the ESO archive
facility. Table~\ref{optobs} shows the log of the optical spectroscopic
observations, summarising the instrumental set-up and giving the equivalent
width of the H$\alpha$ and H$\beta$ lines. The EMMI observation corresponds
to that reported by \citet{negueruela06}, while the EFOSC observations are
unpublished: programs IDs 079.D-0371(A) and 083.D-0110(A). Data reduction
was performed using the IRAF packages, following the standard procedure.


\section{X-ray observations and data analysis}

We analysed data obtained by all three instruments aboard {\em RXTE}
\citep{bradt93}:  the All sky Monitor (ASM) data consist of daily flux
averages in the energy range 1.3-12.1 keV. The Proportional Counter Array
(PCA) covers the lower part of the energy range 2--60 keV, and  consists of
five identical coaligned gas-filled proportional units giving a total
collecting area of 6500 cm$^{-2}$ and provides an energy resolution of 18\%
at 6 keV. The High Energy Timing Experiment (HEXTE) is constituted by 2
clusters of 4 NaI/CsI scintillation counters, with a total collecting area
of 2 $\times$ 800 cm$^2$, sensitive in the 15--250 keV band with a nominal
energy resolution of 15\% at 60 keV. Data taken during satellite slews,
passage through the South Atlantic Anomaly and Earth occultation were
removed. The overall on-source time analysed amounts to 496.3 ks.

The mission-specific packages of Heasarc FTOOLS (version 6.6.3) were
employed to perform data reduction, while XSPEC
v12.6\footnote{http://heasarc.gsfc.nasa.gov/docs/xanadu/xspec/}
was used for spectral analysis. For each observation we
obtained an average energy spectrum and a power spectrum. 

The energy spectra were generated from standard mode data, namely, the PCA
{\em Standard 2} of PCU2 and HEXTE "FS58" cluster B data, which have a time
resolution of 16 s and cover the 2-60 keV range with 129 channels and the
15--250 keV with 64 channels, respectively. All spectra were
background-subtracted.  A systematic error of 0.6\% was added in quadrature
to the PCA spectra to account for systematic errors. The power spectra were
generated by computing Fourier transforms of 128-s segments of the
$2^{-6}$-s binned light curves in the energy range 2--15 keV (PCA channels
0--35) and averaging them together. The final power spectra were
logarithmically rebinned in frequency and corrected for dead time effects
according to the prescription given in \citet{nowak99}. Power spectra were
normalized such that the integral over a given frequency range equals to
the squared fractional $rms$ amplitude \citep{belloni90,miyamoto91}.

To fit the energy spectra, we used a model composed by a combination of
photoelectric absorption, a power law with high-energy exponential cutoff,
a Gaussian line profile at  6.5 keV and an absorption edge at around 9
keV to account for Fe K  fluorescence. We did not find evidence for a
cyclotron absorption feature, commonly seen in other accreting X-ray
pulsars. The power spectra were fitted using Lorentzian profiles only. The
main peak of the X-ray pulsations and two to three of its harmonics are
clearly seen in the power spectra. These peaks were fitted with narrow
Lorentzians (the width was normally fixed at $\sim 0.001$ Hz). The
broad-band noise was fitted with two zero-centred Lorentzians. The brighter
observations required one extra Lorentzian to account for the noise above
$\sim 7$ Hz.

   \begin{figure*}
   \centering
   \includegraphics[width=14cm]{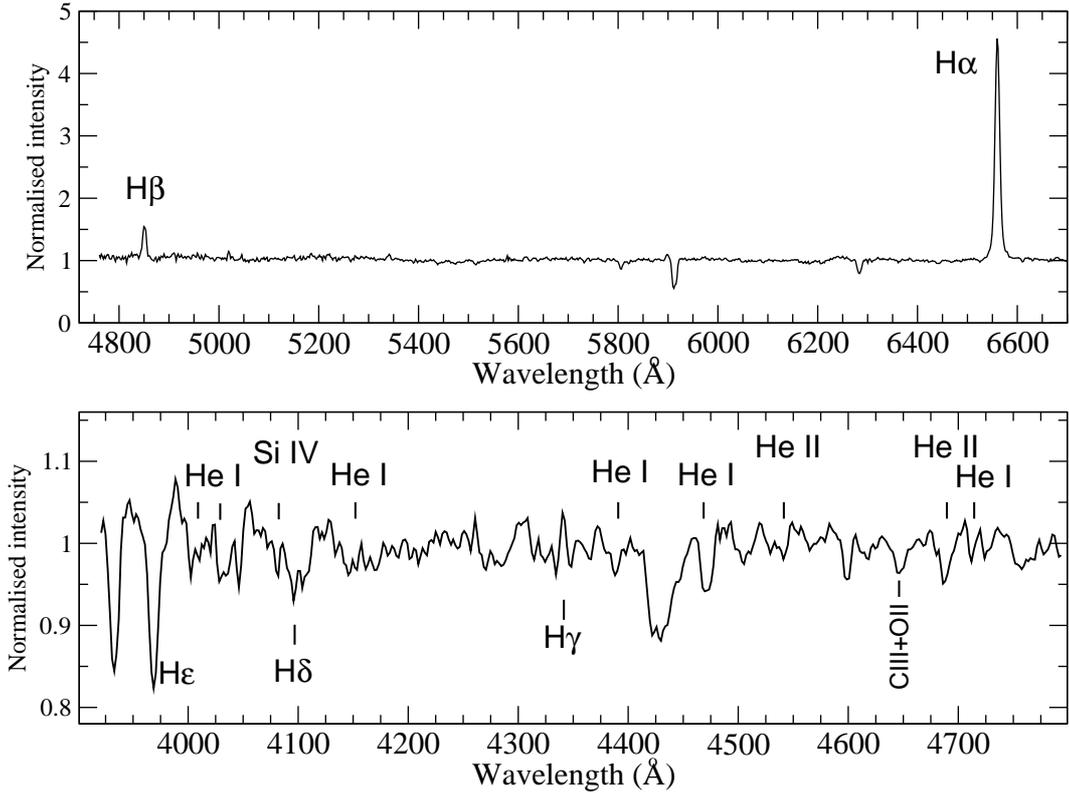} \\
      \caption{Optical spectrum in the classification region. The
      identified lines are He\,I $\lambda$4009,
$\lambda$4026, $\lambda$4144, $\lambda$4387, $\lambda$4471, $\lambda$4713
\AA; He\,II $\lambda$4541 and $\lambda$4686 \AA; Si\, IV $\lambda$4089 \AA; the
C\,III+O\,II blend $\lambda$4650 \AA.}
         \label{optsp}
   \end{figure*}

\section{The optical counterpart}

\subsection{Spectral type}

The information on the optical counterpart to \src\ is very scarce. To the
authors' knowledge no optical photometric observations are published.
The only dedicated observations are those reported in
the Astronomer's Telegrams ATel\#713 \citep{rea06} who obtained infrared
photometric magnitudes and ATel\#739 \citep{negueruela06}, who performed
spectroscopic observations and reported the presence of H$\alpha$ strongly
in emission. 

Information on photometric magnitudes and colours can be found in
astronomical catalogues. We searched for possible optical/infrared
counterparts in the NOMAD\footnote{The NOMAD catalogue is a facility
provided by the U. S. Naval Observatory and contains astrometric and
photometric data for about 1.1 billion stars derived from the Hipparcos,
Tycho-2, UCAC2, Yellow-Blue 6, and USNO-B catalogs for astrometry and
optical photometry, supplemented by 2MASS near-infrared photometry. The
primary aim of NOMAD is to help users retrieve the best currently available
astrometric data for any star in the sky by providing these data in one
place.} catalogue around the best-fit X-ray position \citep{campana06}. 
The optical counterpart to \src\ is the star named as NOMAD1-0380-0705569
(=2MASS16263652-5156305 = USNO-B1.0 0380-0649488) and it is shown in
Fig.~\ref{chart}. The selected star is the only one exhibiting
significant infrared  \citep{rea06} and H$\alpha$ \citep{negueruela06}
emission within the {\it SWIFT}/XRT error circle or its vicinity, which are
the signatures of the vast majority of optical counterparts in accreting
X-ray pulsars. Table~\ref{cat} gives the catalogued photometric
magnitudes.



The low-resolution $K$-band spectrum is almost featureless, except for the
presence of He\,I $\lambda$20\,581 \AA, slightly affected by a residual
telluric absorption and H\,I $\lambda$21\,660 \AA\ (Brackett-$\gamma$, or
Br$\gamma$), which appear strongly in emission. These two lines are typical
of Be-stars and confine the spectral type of the companion star to be
earlier than B2.5 \citep{clark00}. The $H$-band spectrum is more complex,
with the H\,I Br-18-11 recombination series in emission. As for the $K$ band,
all the observed features are typical of Be stars, while they are absent,
or observed in absorption in normal (non-emission) OB stars. The observed
features, with measured equivalent widths are shown in Table~\ref{ewir}. 

By comparison with spectral atlases \citep{clark00, steele01}, the optical
counterpart to \src\ can be classified as a B0-2Ve star, confirming the
nature of the system as a high-mass X-ray binary. Note that  all
spectroscopically identified optical companions of Be/X-ray binaries in the
Milky Way have spectral types in the narrow range O9-B2, with a peak around
B0 \citep[see e.g. Fig.6 in][]{antoniou09}.

Based on $JHK$ photometry, \citet{rea06} suggested that the optical
counterpart to \src\ could not be an early-type star because after
de-reddening using the X-ray hydrogen column density ($N_{\rm H}=0.9\times
10^{21}$ cm$^{-2}$), the infrared magnitudes and colours were too faint to
be consistent with an OB star.  The value of $N_{\rm H}$ corresponded to a
{\it Swift}/XRT observations reported by \citet{campana06}.  However,  the
actual value of $N_{\rm H}$ reported in \citet{campana06} is an order of
magnitude larger than that used by \citet{rea06}. We repeated the
computation of the intrinsic colour of the source, employing the {\it
SWIFT}/XRT $N_{\rm H}$ value from \citet{campana06}, $N_{\rm
H}=(9.4\pm1.0)\times 10^{21}$ cm$^{-2}$, which is more reliable than our
{\it RXTE} measurement due to the softer energy range sensitivity. This implies a
visual interstellar extinction  $A_V=4.25\pm0.45$ mag \citep{guver09},
which converts into an interstellar IR color excess
$E^{is}(H-K)=0.063 \times A_V
= 0.27\pm0.03$ mag \citep{rieke85}. Note that the dense circumstellar
material surrounding the Be star introduces extra reddening that cannot be
ignored, especially at long wavelengths \citep{dachs88,dougherty94}. Hence
$E^{tot}(H-K)=E^{is}(H-K)+E^{cs}(H-K)$. The reddening caused by the disc
can be estimated from the relationship between infrared excess and the
H$\alpha$ equivalent width from \citet{howells01}, who derived
$E^{cs}(H-K)=0.006EW(H\alpha)-0.030$ mag. We do not know the state of the
disc when the infrared observations were measured, but if we assume that
the H$\alpha$ equivalent width was at the level reported here $\sim -40$
\AA, then $E^{cs}(H-K)=0.21\pm0.02$ mag and  the resulting intrinsic colour is
$(H-K)_0 = (H-K)_{\rm 2MASS} - E^{tot}(H-K)=-(0.07\pm0.04)$ mag, consistent
with an early-type star \citep{koornneef83}. The errors were obtained by
propagating the errors of the measurements.

The September 2007 blue-end spectra allow us to refine the spectral
type estimated from the near-IR data. Figure~\ref{optsp} shows the optical
spectrum of \src\ in the traditional classification region (3800--4800
\AA). Note that emission affects the lines of the Balmer series: H$\beta$
and H$\gamma$ are in emission, while H$\delta$ is partially filled-in with
emission. The spectrum contains many He\,I lines ($\lambda$4009,
$\lambda$4026, $\lambda$4144, $\lambda$4387, $\lambda$4471, $\lambda$4713
\AA), indicating an early B-type star. The presence of He\,II
($\lambda$4541 and $\lambda$4686 \AA) indicates that the spectral type of
the optical star is earlier than B1. Specifically, He\,II $\lambda$4686
\AA\ is last seen at B0.5-B0.7 \citep{walborn90}. On the other hand, the
weakness of He\,II $\lambda$4541 \AA\ relative to He\,I $\lambda$4471 \AA\
indicates an spectral type later than O9.

With regard to the luminosity classification, the weakness of Si\,III 
$\lambda$4552--68 \AA\ and the fact that the line intensity of He\,II
$\lambda$4686 \AA\ is larger than that of He\,I $\lambda$4713 \AA\ seem to
indicate a main-sequence star. Other luminosity indicators are the ratios
of He lines to nearby metallic lines. \src\ shows He\,II $\lambda$4686/CIII
$\lambda$4650 $\simmore$ 1 and He\,I $\lambda$4144/Si\,IV 4089 $\sim$ 1. An
evolved star would have those ratios $<$ 1 and $\simless$ 1, respectively.

A visual comparison of the \src\ spectrum (Fig.~\ref{optsp}) with those of
MK standards in the atlas by \citep{walborn90} reveals that the spectrum
resembles that of the standard star $\upsilon$ Ori, a B0V star. Therefore,
we conclude that the optical counterpart to the X-ray accreting pulsar
\src\ is a B0Ve star.

   \begin{figure}
   \centering
   \includegraphics[width=8cm]{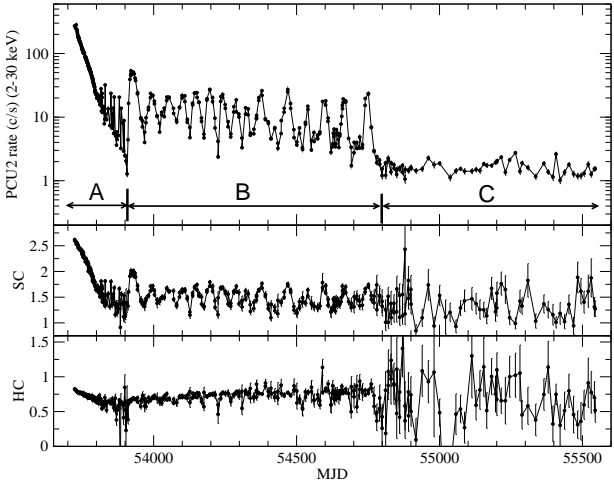}
   \caption{X-ray light curve of \src\ showing five years worth of data. 
   Each data point corresponds to the average count rate of the
   individual pointings. Also shown are the soft colour, SC=4-7 keV/2-4 keV
   and hard colour, HC=10-15 keV/7-10 keV.}
              \label{lc_hr}
    \end{figure}
   \begin{figure}
   \centering
   \includegraphics[width=8cm]{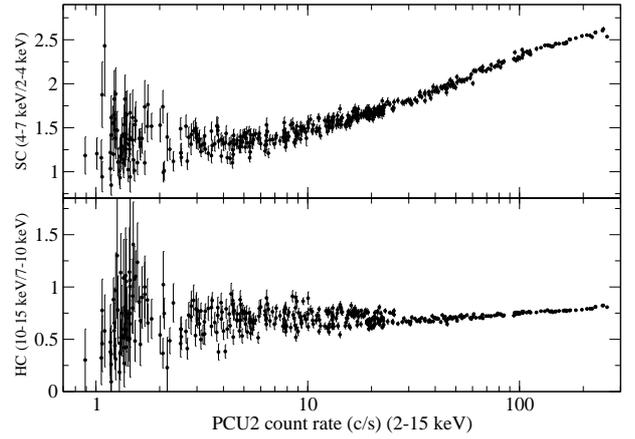}
   \caption{Hardness-intensity diagram.}
              \label{hid}
    \end{figure}

\subsection{Distance}
\label{dist}

Although the uncertainty of the spectral type may introduce some extra
uncertainty in the distance estimation, the main source of error in
estimating the distance stems from the scattering of the intrinsic colours
and absolute magnitudes associated to each spectral and luminosity class of
the calibrations. Even for the same spectral type, variations of up to 1.5
mag in the absolute magnitude \citep{wegner06} and of 0.04 mag in the
intrinsic $(B-V)_0$ colours are found in studies from various authors
\citep{johnson66,wegner94}. \citet{jaschek98} analysed the absolute
magnitude of about 100 MK standards and concluded that the intrinsic
dispersion of the mean absolute magnitude amounts to 0.7 mag. We have
adopted this value as the error on the absolute magnitude.

The BVRI magnitudes listed in Table~\ref{cat} are photographic magnitudes,
extracted from the USNO-B1 and the YB6 surveys, and hence they are not
directly comparable with the intrinsic colours and absolute magnitude
calibrations of the Johnson-Cousins photometric system. This prevent us
from using these magnitudes to compute the interstellar reddening and the
distance to our target. We will approach this task by using the 2MASS JHK
magnitudes also listed in Table~\ref{cat}.

Assuming that the optical counterpart to \src\ is a B0V star, then
$(J-K)_0=-0.17$ mag \citep{koornneef83}. The total IR excess is
$E^{tot}(J-K)=1.07$. Part of this excess is due to the circumstellar
continuum emission of the Be star. We have calculated the circumstellar
contribution to the  IR excess to Be $E^{cs}(J-K)=0.25$ mag from the
equivalent width of the H$\alpha$ line and the H- and K-band spectral
features, by means of the formulae presented by  \citet{howells01}. Thus we
obtained the interstellar IR excess as
$E^{is}(J-K)=E^{tot}(J-K)-E^{cs}(J-K) = 0.82$ mag. This implies a visual
extinction of $A_V=4.7$ mag \citep{koornneef83}, little higher, but consistent
with the value obtained from the X-ray observations. The absolute magnitude
of a B0V star is $M_V=-3.95$ mag  \citep{wegner06}, which can be converted
into $M_K$ since $(V-K)_0=-0.85$ mag \citep{koornneef83}; the intrinsic
magnitude $K_0$ was computed as the difference of the 2MASS magnitude and
the K-band extinction $A_K$ resulting from the relation
$A_\lambda/E(J-K)=2.4(\lambda)^{-1.75}$, for $\lambda = 2.2 \mu$m
\citep{draine89}. We finally calculated the distance from 
$M_K=K_0+5-5logd$, obtaining $d=10.7\pm3.5$ kpc. This value must be
considered as a lower limit, since we neglected the circumstellar
contribution to $M_K$. The error was estimated assuming an error in
the observed infrared magnitudes (hence $E(J-K)$) of 0.03 mag and 0.7 mag
in the absolute magnitude. 


\section{Long-term X-ray variability}

The long-term X-ray light curve of \src, covering five years worth of data,
since its discovery on 19 December 2005 (MJD 53723) up to 14 December 2010
(MJD 55544), is shown in  Fig.~\ref{lc_hr}. This figure also shows the
evolution of the X-ray colours defined by the ratio of the count rates in
the energy range 4-7 keV over 2-4 keV (soft colour, SC) and 10-15 keV over
7-10 keV (hard colour, HC). The X-ray flux of \src\ decreased exponentially
during the first 90 days of the observations. On 20 March 2006 (MJD 53814)
the source began to exhibit low-amplitude X-ray flares whose duration
varied between 3-8 days. Nevertheless the overall decrease in flux
continued until 26 June 2006 (MJD 53912), where a large flare took place.
After the large flare, the mean intensity increased to $\sim$10 c s$^{-1}$
PCU$^{-1}$ and the X-ray flaring behaviour became more apparent. A total of
15 flares can be seen in Fig~\ref{lc_hr}. The time difference between the
peak of the flares varied from $\sim$45 days of the first half to $\sim$95
days of the second half \citep{reig08,baykal10}.

On 29 November 2008 (MJD 54799) the X-ray count rate decreased suddenly by
almost an order of magnitude to $\sim$ 1.5 c s$^{-1}$ PCU$^{-1}$ and the
large-amplitude flaring pattern ceased.  The source entered a quiescent
phase.
Note that the source was observed daily until MJD 53827, every 2-4 days
until MJD 53936, every 9-10 days until MJD 54900 and in a more irregular
way every 15-20 days (but also including a few closer observations) until
the end of the observations on MJD 55544.

The soft colour follows closely the evolution of the X-ray flux
(Fig.~\ref{hid}). As the count rate increases so does the SC. The HC also
shows a positive correlation with flux. However, this is weaker than that at
lower energies and it is evident only at higher count rates $\simmore 20$ c
s$^{-1}$ PCU$^{-1}$. The correlation seems to extend to the lower count
rate data but the scattering of the data points due to poor statistics is
too large  below $\sim$ 3 c s$^{-1}$ PCU$^{-1}$.

Given the richness in X-ray variability exhibited by \src, we have defined
three different time intervals and performed a separated analysis on each
one. These periods correspond to the smooth decay part of the outburst,
including the small-scale flaring period (MJD 53723--53912),  the large
scale flaring (MJD 53912--54799) and the quiescent interval (MJD
54799--55544). 

   \begin{figure}
   \centering
   \includegraphics[width=8cm]{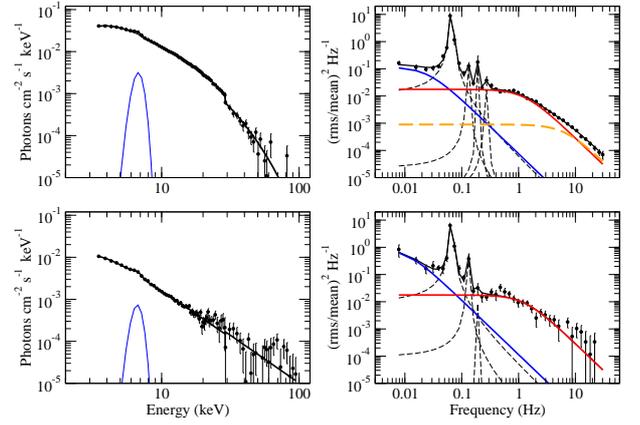}
   \caption{Energy and power spectra at high (upper panels) and low (lower
   panels) X-ray flux. The X-ray flux is $\sim 4 \times 10^{-9}$ and
   $\sim 5 \times 10^{-10}$erg cm$^{-2}$ s$^{-1}$, respectively.}
    \label{spec-psd}
    \end{figure}

   \begin{figure*}
   \centering
   \includegraphics[width=15cm]{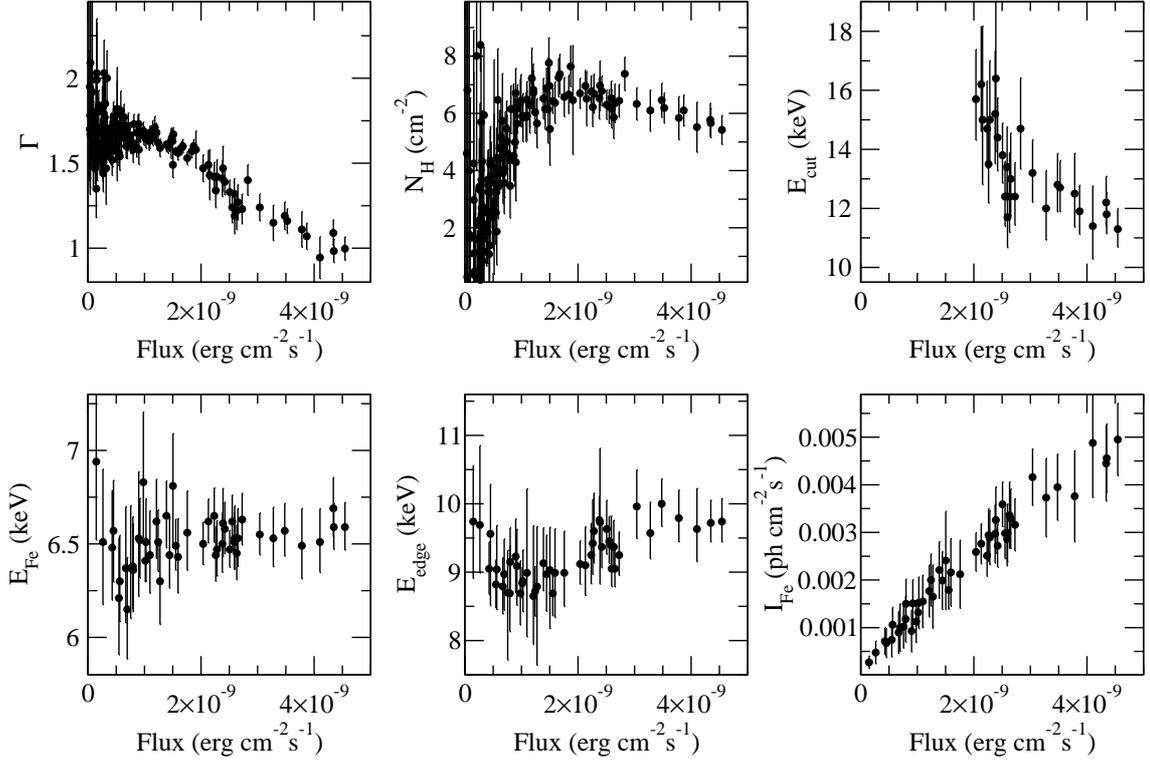}
      \caption{Evolution of the spectral parameters over the decay of the
      outburst. Flux corresponds to the 3-30 keV energy range. 
      Power-law index, hydrogen column density and cutoff energy ({\em top
      panels}) and central energy, edge energy and intensity of the iron
      line ({\em bottom panels}) as a function of X-ray flux.}
         \label{specpar_decay}
   \end{figure*}

   \begin{table*}
   \begin{center}
      \caption[]{Average best-fit spectral parameters during the
      large-amplitude flares and the quiescent state that followed.}
         \label{flaring_spec} 
         \begin{tabular}{lcccc|c}
            \hline  \noalign{\smallskip}
	&\multicolumn{4}{c}{Flaring state}	& Quiescent state \\
Parameter	&Minimum	&rise	&maximum	&decay \\
            \noalign{\smallskip}  \hline
            \noalign{\smallskip}
Photon index ($\Gamma$)			&$1.9\pm0.1$	&$1.5\pm0.1$	&$1.16\pm0.09$	&$1.6\pm0.1$	&$2.10\pm0.08$   \\
Cutoff energy$^{*}$ ($E_{\rm cut}$) 	&$30^f$		&$20^{+5}_{-3}$	&$11\pm1$&$21^{+8}_{-5}$	&-- \\
Normalization$^{**}$ 			&$0.017\pm0.001$&$0.031\pm0.003$&$0.034\pm0.003$&$0.032\pm0.004$&$0.007\pm0.001$ \\
Line energy$^{*}$ ($E_{\rm Fe}$)	&$6.9\pm0.1$	&$6.7\pm0.1$	&$6.6\pm0.1$	&$6.7\pm0.1$	&$6.6\pm0.1$ \\
Line width$^{*}$ ($\sigma_{\rm Fe}$)	&$0.4^f$	&$0.3^f$	&$0.2^f$	&$0.3^f$	&$0.1^f$\\
Edge energy$^{*}$ ($E_{\rm edge}$)	&$9^f$		&$8.5\pm0.2$	&$8.3\pm0.1$	&$8.5\pm0.3$	&--	\\
Edge optical depth ($\tau_{\rm edge}$)	&$0.07^{+0.10}_{-0.07}$	&$0.17\pm0.05$	&$0.26\pm0.03$	&$0.17\pm0.06$&--	\\
$\chi^2$/dof				&1.9/48		&1.6/46		&1.1/46		&1.2/45		&1.0/36	\\
            \noalign{\smallskip}
            \hline
         \end{tabular}
\tablefoot{$^*$: keV; $^{**}$: ph keV$^{-1}$cm$^{-2}$ s$^{-1}$ at 1 keV; $^f$: fixed}
	\end{center}
   \end{table*}

\subsection{Outburst decay (MJD 53723--53912)}

\src\ was discovered on 18 December 2005 when the source was near the peak
of a major outburst. The X-ray flux of the first observations was $4.5
\times 10^{-9}$ erg cm$^{-2}$ s$^{-1}$ and decreased for the next $\sim$190
days (interval A in Fig.~\ref{lc_hr}). Figure~\ref{spec-psd} shows two 
representative energy and power spectra of the outburst decay at fluxes
differing by about one order of magnitude. The spectral parameters changed
smoothly as the outburst decayed (Fig.~\ref{specpar_decay}). However, at
flux below $F_{\rm break} \sim 1.5 \times 10^{-9}$ erg cm$^{-2}$ s$^{-1}$
the source experienced a significant change in its spectral shape. The
X-ray continuum of the brighter observations, i.e., when the 3-30 keV flux
was  above $F_{\rm break}$, showed  a
faster decay at high energies requiring  an exponential cutoff to fit the
data. The energy of the cutoff increased as the flux decreased. Below
$F_{\rm break}$ this component was not needed and the X-ray continuum was
well represented by a single absorbed power law extending to energies up to
100 keV (Fig.~\ref{spec-psd}). Likewise, the hydrogen column density,
$N_{\rm H}$ fell rapidly  as the flux decreased below $F_{\rm break}$,
while it remained at a constant value of $\sim 6.5 \times 10^{22}$
cm$^{-2}$ above $F_{\rm break}$. The photon index decreased as the flux
increased, that is, when the source was bright the X-ray emission was
harder. At low flux, roughly coincident with $F_{\rm break}$ the photon
index  flattened at a value of 1.7 (Fig.~\ref{specpar_decay}).  Near the
peak of the outburst the photon index was $\sim$ 1.1. 

The presence of the iron emission line or the absorption edge was not
always statistically significant. When the line parameters could not be
well constrained, the line and/or edge energy were fixed at their average
values, prior to the fit. The mean value of the central  and absorption
edge energies were calculated using observations in which the intensity of
the line (normalization) was more than $2\sigma$ significant and resulted
in $E_{\rm line}=6.5\pm0.1$ keV and $E_{\rm edge}=9.3\pm0.4$ keV, where the
errors represent the standard deviation of all the measurements. The three
bottom panels of Fig.~\ref{specpar_decay} show the dependence of the iron
line energy and intensity and the edge energy with the 3-30 keV X-ray
continuum flux. There is a tight correlation between the X-ray continuum
and the intensity of the iron line, indicating that as the illumination of
the cool matter responsible for the line emission increases, so does the
strength of the line.

The characteristic frequency of the broad-band noise increased as the flux
increased, but saturated at $\nu\approx 1.5$ Hz above $F_{\rm break}$. The
large error bars at low flux weakens the statistical significance of this
trend. Almost identical trend was followed by the fractional amplitude of
variability. Above $\sim 2.6 \times 10^{-9}$ erg cm$^{-2}$ s$^{-1}$ there
is excess power at high frequencies and an extra noise component with
characteristic frequency at $\sim 6-7$ Hz and $rms$ 10-15\% was needed to
obtain acceptable fits (upper right panel in Fig.~\ref{spec-psd}). Some of
the brightest observations contain a weak QPO with centroid frequency at
$\sim$ 1 Hz and $rms \sim 3$\%. Although the inclussion of this component
improves the fit, an F-test shows that the probability that this
improvement occurs by chance is $\sim$1--3\%.

\subsection{Large-scale flaring emission (MJD 53912--54799)}

At the end of the outburst decay and after a series of low-amplitude
flares, \src\ began to display quasiperiodic increases in the X-ray
intensity (interval B in Fig~\ref{lc_hr}). The maxima and minima of these
flares remained roughly at the same flux level, with $F_{\rm max}=(28\pm5)
\times 10^{-11}$ erg cm$^{-2}$ s$^{-1}$ and $F_{\rm min}=(5\pm2) \times
10^{-11}$ erg cm$^{-2}$ s$^{-1}$. The only exception was the first flare,
which started  and reached $\sim$2.5 times lower and higher flux than the
average minimum and maximum flux, respectively. The timescale of the X-ray
modulation varied between 45 and 95 days, although most of the time
separation between flares gathered around $P_{\rm orb}/2$ and $P_{\rm
orb}/3$, being $P_{\rm orb}=133$ days  \citep[see][for an analysis of this
behaviour]{baykal10}.

Due to the low count rate, the energy spectra above $\sim$10 keV and the
power spectra above $\sim$ 1 Hz are too noisy for a meaningful analysis. 
The cutoff energy and especially the iron emission line parameters cannot
be constrained in the individual spectra, which represent data obtained by
integrating over 0.5-1 ks, typically. However, by joining observations with
roughly the same flux, those components become significant. Therefore in
order to increase the signal-to-noise ratio we obtained an averaged peaked,
trough, intermediate-rise and intermediate-decay flux energy and power
spectrum. The total on-source time for these average spectra was 17.6,
19.1, 15.4 and 11.5 ks, respectively.  Table~\ref{flaring_spec} gives the
results of the spectral fits. The spectrum at the peaks is harder than that
of the troughs. The cutoff energy follows the same trend with X-ray flux as
that seen during the outburst decay, namely, it increases as the flux
decreases. The characteristic frequency of the broad-band noise component
also agrees with the values that the source showed during the decay of the
outburst for the same flux level. At the peak of the flares  $\nu_{\rm
max}=0.40\pm0.09$ Hz, while at intermediate flux  $\nu_{\rm rise}\approx
\nu_{\rm decay}=0.12\pm0.05$ Hz. The fractional amplitude of variability is
$\sim$ 25\% in all cases. The timing parameters at the minimum flux of the
flares cannot be constrained due to the low signal-to-noise ratio of the
power spectrum.

\subsection{Quiescence (MJD 54799--55544)}

After the series of flares the source entered a quiescent state,
characterised by little variability. The X-ray flux decreased to an average
level of $(1.7\pm0.4) \times 10^{-11}$ erg cm$^{-2}$ s$^{-1}$. Again, the
individual spectra are too noisy to be able to constrain the spectral
parameters. Neither an iron emission line nor the exponential cutoff are
statistically required in the individual spectra. However, when an average
energy spectrum is obtained by combining all observations in the interval
MJD 54799--55544 then a good fit cannot be obtained unless a Gaussian
component at 6.5 keV is included, although no edge is required. The
inclusion of that component reduced the $\chi^2$ statistic from 94 for 38
degrees of freedom  to 35 for 36 degrees of freedom.  An F-test gives a
probability that the improvement of the fit occurs by chance of only
$1.9\times 10^{-8}$. The total on-source time of this spectrum was 46.4 ks,
corresponding to 61 observations. The best-fit spectral parameters are
given in Table~\ref{flaring_spec}. The energy range was limited to 3-20
keV. The spectrum during the quiescent state is softer than  during the
outburst decay and during the flaring state, but follows the general trend
also observed in the other two states that the lower the flux, the larger
the photon index, i.e., the softer the spectrum.

   \begin{figure}
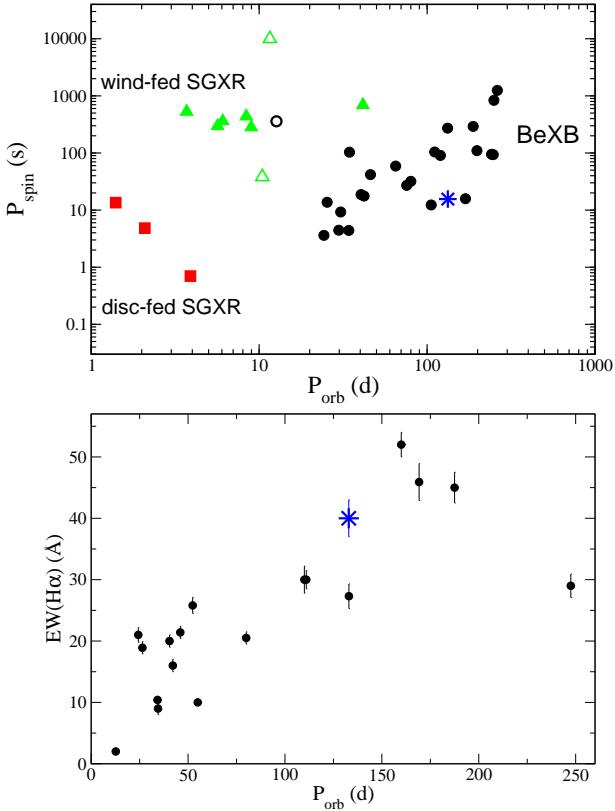

   \centering
   \begin{tabular}{c}
   \includegraphics[width=8cm]{./17301f8a.eps} \\
   \includegraphics[width=8cm]{./17301f8b.eps} \\
   \end{tabular}
      \caption{$P_{\rm orb}-P_{\rm spin}$ and $P_{\rm orb}-EW(H\alpha$)
      diagrams. The star symbol marks the position of \src. Open symbols in
      the top panel correspond to the peculiar SGXBs 2S 0114+65 and OAO
      1657--41 and the BeXB SAX J2103.4+4545.}
         \label{diagrams}
   \end{figure}

\section{Discussion}

The available data in various wavelength bands provide strong evidence that
the system is a Be/X-ray binary. In the X-ray band, the detection of 15.4-s
pulsations and the shape of the spectral continuum (a power law with
exponential cutoff) are characteristic of accreting X-ray pulsars.  In
the optical band, the strong H$\alpha$ emission line and the presence of He
\,I and He\, II lines indicate an OB-type star as the optical counterpart. In
the NIR, both the $H$- and the $K$-band observed features (HeI end HI
recombination series in emission) are also typical of early Be stars. This
allows us to classify the counterpart to \src\ as a B0Ve star, and the
system as a Be/X-ray binary.

Further evidence in favour of a Be/X-ray binary classification
comes from the $P_{\rm orb}-P_{\rm spin}$ diagram \citep{corbet86} and the
$P_{\rm orb}-EW(H\alpha$) diagram \citep{reig97,reig11}.
Figure~\ref{diagrams} shows these two diagrams. As can be seen, \src\ (big
star symbol) agrees with the $P_{\rm orb}-EW(H\alpha)$ correlation and falls
in the Be/X-ray binaries region of the $P_{\rm orb}-P_{\rm spin}$ diagram.

The optical spectra that cover the region of the H$\alpha$ line show
comparable values of the H$\alpha$ equivalent width despite the fact that
they correspond to very different X-ray states and were taken  years apart.
The February 2006 observation corresponds to the middle of the X-ray
outburst decay, the September 2007 observations correspond to the middle of
the flaring episode and the June 2009 observation to the beginning of the
quiescent X-ray state. In Be stars, the strength and shape of the H$\alpha$
line provides information about the physical conditions in the
circumstellar disc. Generally, after a major X-ray outburst, the H$\alpha$
equivalent width decreases substantially indicating a weakening of the disc
\citep{coe94} or even the line shape turns from an emission profile to an
absorption profile, suggesting that the disc is lost \citep{reig07}. 
In contrast, the disc in \src\ does not seem to be affected by the X-ray
outburst, as the H$\alpha$ equivalent width did not vary much immediately
after the X-ray outburst nor a few years later. 


The behaviour of \src\ resembles that of the Be/X-ray binary KS 1947+300.
As \src, KS 1947+300 went through a series of periodic increases of the
X-ray intensity (type I outburst) after a major (type II) outburst (see
Fig. 11 in \citealt{reig11}) without the H$\alpha$ line noticing it.
Another characteristic in common with KS 1947+300 is the near-circular
orbit. The eccentricity in these two systems (0.03 and 0.08, respectively)
are among the smallest in Be/X-ray binaries. An alternative scenario would
be that of 1A 1118-61, which exhibited a huge equivalent width ($-90$ \AA)
before the X-ray outburst \citep{coe94}. After the outburst, the H$\alpha$
equivalent width decreased substantially but remained at a relatively high
level around $-60$ \AA\ for the following years \citep{villada99}.

A discrete component at $\sim$6.5 keV is always present, albeit weak, in
the X-ray energy spectra of \src\ in all states. This component is interpreted as
reprocessing of the hard X-ray continuum in relatively cool matter. Near
neutral iron generates a line centered at 6.4 keV, while this energy
increases as the ionisation stage increases. However, the energy separation
is so small that even ionised iron up to Fe XVIII can be thought as part of
the 6.4 keV blend \citep{liedahl05}. The line energy during the outburst
decay is consistent with this neutral iron blend. The line energy during
the flaring state seems to be larger. However, given the weakness of the
line we do not claim a higher ionisation degree in this state. What it is
important to notice here is that the fluorescent iron line provides strong
evidence for the presence of material in the vicinity of the X-ray source.
It has been detected in virtually all high-mass X-ray binaries with
supergiant companions. In these systems the Fe line fluorescence is
produced in the stellar wind of the massive star \citep{torrejon10}. The
possible site for reprocessed emission in a Be/X-ray binary are an
accretion disc or the circumstellar (decretion) disc. The former surrounds
the neutron star, while the latter surrounds the Be star's equator. The
detection of this component in the energy spectrum of the quiescent state
would favour the equatorial disc as the site for the reprocessing of
hard-energy photons, as we would expect that the accretion disc, if ever
present, would have disappeared. 
The alternative scenario that the Fe iron line emission is produced in the
thin thermal hot plasma that is presumably located along the Galactic plane
\citep{yamauchi09} is ruled out by the observed correlation between the
line intensity and the X-ray continuum flux.

We estimate the distance to be $\sim$10 kpc with an error of the order of
30\%. This distance and the Galactic coordinates of the source ($l=332.8,
b=-2.0$), place the system in the Norma Arm of the Galaxy. The sources of
error in the determination of the distance can be attributed to: {\em i)}
uncertainties in the calibrations that give intrinsic colours and absolute
magnitudes as a function of spectral types, {\em ii)} the presence of a
large circumstellar disc which contaminates the optical and IR colours
making the observed colours appear redder that their intrinsic values, and
to a lesser extent, {\em iii)} the uncertainty in the spectral type. 

Note that in Be stars the total reddening not only includes the
contribution of the interstellar extinction but also of the reddening due
to the circumstellar material. The contribution of the circumstellar disc
to the photometric colours increases with wavelength from $\sim$0.1 mag in
$(B-V)$ to $\sim$0.8 mag in $(J-M)$ for well-developed discs
\citep{dachs88}. Moreover, it has been shown that the circumstellar
reddening correlates with the equivalent width of the H$\alpha$ line
\citep{dachs88,fabregat98,howells01}. The larger the equivalent width, the
larger the contribution of the circumstellar emission to the total
reddening.  Since \src\ shows strong H$\alpha$ emission with an equivalent
width of --45 \AA, the disc emission can make an important contribution to
the colour excess. Although we try to account for this extra extinction
using the equations by \citet{howells01}, it is worth noticing that the
photometric infrared magnitudes and the optical and infrared spectra were
not contemporaneous, which introduces extra uncertainty in the distance
estimation given the high-amplitude variability that characterises the
optical and infrared emission of Be stars. Our distance estimate is
 consistent with the value of $\sim$15 kpc reported by
\citet{icdem11}. These authors estimated the distance from the correlation
between spin-up rate and X-ray flux. Although \citet{icdem11} do not give
the error of their measurement, the application of this method in other
sources has shown that the uncertainty is of the order of $\sim$10--30\%.


\section{Conclusion}

We have solved the nature of the X-ray pulsar \src. The new optical
and infrared data provide strong evidence that the system is a Be/X-ray
binary. The location of \src\ in the $P_{\rm orb}-P_{\rm spin}$ and
$P_{\rm orb}-EW(H\alpha)$ diagrams, the X-ray pulsations, the shape of the
X-ray continuum, the long-term X-ray variability, the emission of the
Balmer and Brackett lines in the optical/IR spectrum are typical
characteristics of high-mass X-ray binaries with a Be companion. We
conclude that \src\ is a Be/X-ray binary with a B0Ve optical counterpart
located at a distance of $\sim$10 kpc. Optical photometric observations are
needed to reduce the uncertainty in the distance estimation.

\begin{acknowledgements}

We thank the referee, Malcolm Coe, for his useful and important comments
that improved the final version of this manuscript. The work of EN y JF is
supported by the Spanish Ministerio de Educación y Ciencia, and FEDER,
under contract AYA2010-18352. PR and JF are partially supported by the
Generalitat Valenciana project of excellence PROMETEO/2009/064. PR also
acknowledges partial support by the COST Action
ECOST-STSM-MP0905-280611-008632. REM acknowledges support by Fondecyt grant
1070705, 1110347, the Chilean  Center for Astrophysics FONDAP 15010003 and 
from the BASAL Centro de Astrof\'isica y Tecnologias Afines (CATA)
PFB--06/2007. This work was based on observations collected at the European
Southern Observatory, Chile (Programme ID: 085.D-0297A) and on data
obtained from the ESO Science Archive Facility. This research has made use
of the USNOFS Image and Catalogue Archive operated by the United States
Naval Observatory, Flagstaff Station
(http://www.nofs.navy.mil/data/fchpix/).

\end{acknowledgements}

\end{document}